\documentclass{article}
\usepackage{graphicx} 
\usepackage{amsmath}
 \usepackage{natbib}
\usepackage{newunicodechar}
\usepackage{mathrsfs} 
\DeclareTextCommandDefault{\textyen}{%
  \vphantom{Y}%
  {\ooalign{Y\cr\hidewidth\yenbars\hidewidth\cr}}%
}

\newcommand{\yenbars}{%
  \vbox{
     \hrule height.1ex width.4em
     \kern.15ex
     \hrule height.1ex width.4em
     \kern.3ex
  }%
}

\title{What is the connection between quantum relationality and symmetry-based relationality?} 
\date{\today}
\author{Emily Adlam}

\begin{document}

\maketitle

\begin{abstract}
  Although it has often been noted that both quantum mechanics and relativistic physics involve significant forms of relationality, the nature of the connection remains unclear. In addressing such a  question it makes sense to start from the simplest possibility, so  in this article I set out to consider the viability of the hypothesis that the two types of relationality are ultimately the same: that is, quantum states always characterize degrees of freedom which can be changed by a symmetry transformation, so they are relativized for exactly the same reason as partial observables in general relativity.  I  use the idea that we only measure partial observables to argue that this  suggestion is compatible with all existing empirical evidence for quantum mechanics, and then I consider the consequences  for ongoing debates about the nature of measurement in quantum mechanics. 
    
\end{abstract}

\section{Introduction}

Relational and perspectival solutions to the measurement problem have attracted significant attention recently \citep{1996cr, Morganti,https://doi.org/10.48550/arxiv.1801.09307,pittphilsci20397,pittphilsci19664}. It has sometimes been suggested \citep{1996cr, Oldofredi_2021,vidotto2022relationalontologycontemporaryphysics} that the fact that we see  relational  features in both quantum mechanics and relativity is significant, perhaps indicating that relational interpretations of quantum mechanics must be on the right track.

Against this, one might object that the word `relational' is so very broad and general that almost anything can be described as relational in some sense, so it is not clear that the mere existence of relational features in both theories is a meaningful signal. Moreover many existing   relational approaches to quantum mechanics appear quite far removed from the type of relationality that appears in relativistic physics. In particular, relativistic relationality is strongly linked with symmetries - it arises from the fact that `partial observables' which can be altered by a symmetry transformation  must be relativized to a reference system in order to be meaningful \citep{Rovelli_2002}. By contrast, the central argument for relationality in quantum mechanics comes from the Wigner's Friend scenario \citep{Wigner}, which appears to have nothing to do with symmetry.  So in order for arguments based on the existence of relationality in both contexts to be compelling, we need a better understanding of the connection between them.

In any such conceptual enquiry it is logical to begin by exploring the simplest possibility, which in this case is that the two types of relationality are ultimately the same.  Thus in this article I set out to consider the viability of the  hypothesis that   quantum states always characterize  degrees of freedom which can be changed by a symmetry transformation, so they must be relativized to a reference system for exactly the same reason that partial observables need to be relativized to a reference system. I will use the idea that we only measure partial observables to argue that this   suggestion is compatible with all existing empirical evidence for quantum mechanics, and I will consider the consequences  for ongoing debates over the nature of measurement in quantum mechanics.

I  begin in section \ref{relational} by describing the principal motivations for relativization in quantum mechanics and relativistic physics. Following \cite{Rovelli_2002, Rovelli_2014}, I argue that we can only measure partial observables. Then in section \ref{connect} I discuss  quantum states of symmetry-variant properties, arguing that in this case the relational nature of the states can be straightforwardly explained as a consequence of the fact that symmetry-variant properties need to be relativized to a suitable reference system. In section \ref{variant} I argue that if it is accepted that we  only measure partial observables, this  explanation of quantum relationality may be extended to all quantum states that can be experimentally demonstrated. I then appeal to a framework developed by \cite{Loveridge_2018} which shows that in a quantum theory with an appropriate symmetry group, we can begin with a state of a composite system which does not exhibit any coherence and then derive from it a state of one subsystem relative to the other which \emph{does} exhibit coherence, thus demonstrating that it is in principle consistent to maintain that partial observables are quantum degrees of freedom while the corresponding complete observables are not. 

Finally, in section \ref{measurement} I discuss some consequences that would follow from the idea that all quantum degrees of freedom are  partial observables. In particular, this approach provides a natural explanation for the fact that we never  obtain superpositions of different measurement outcomes, and suggests new ways of understanding  what is so physically distinctive about quantum measurement and also when and how a  `collapse' occurs during the process of measurement.

\section{Relativization in Quantum Mechanics and Relativity \label{relational}} 

In this section I will briefly set out the central motivations  for relativization in quantum mechanics and relativity, before going on  in section \ref{partial}  to introduce the idea that we only measure partial observables. 

\subsection{Relativized States in Quantum Mechanics \label{quantum}}

The most significant argument for relativization in quantum mechanics comes from the  Wigner's Friend scenario \citep{Wigner} and variations on it \citep{Bong_2020,Cavalcanti_2021,ormrod2022nogo}. Consider an observer, Alice, performing a measurement in the basis $\{ | \uparrow \rangle \langle \uparrow |_S, |\downarrow \rangle \langle \downarrow |_S \}$ on a two-dimensional system $S$  originally prepared in the state $\psi_S = x | \uparrow \rangle_S + y | \downarrow \rangle_S$, with both $x$ and $y$ non-zero. Presumably Alice will see an  outcome to her measurement,  and the standard textbook formulation of quantum mechanics tells us that if Alice sees the outcome $0$ the state of system $S$ is now $| 0 \rangle$, while if Alice sees the outcome $1$ the  state of system $S$ is now $| 1 \rangle$. 

But if another observer, Bob, is observing from the outside and describing the entire interaction using unitary quantum mechanics, he will predict that  $S$ and Alice are now in the following entangled superposition state: 

\[ \psi_{SA} = x | \uparrow \rangle_S | see \ up  \rangle_A + y | \downarrow \rangle_S | see \ down \rangle_A \]

Here, $| see \ up \rangle_A$ and $| see \ down \rangle_A$ are states of Alice corresponding to `seeing the up outcome' and `seeing the down outcome' respectively. This scenario appears to present us with a contradiction:  Alice has surely experienced a single outcome, and yet the state assigned by Bob appears to say that Alice is somehow  experiencing two different outcomes. 

One  notable feature of the Wigner's Friend scenario is that Alice and Bob are now assigning different states to one and the same system at the same time.   For example, suppose   Alice obtained outcome `up' for her measurement - in that case  she now assigns to $S$ the pure state $ | \uparrow \rangle_S$. Whereas 
 Bob, regardless of Alice's outcome, will always assign to $S$ the mixed state $\rho_S = |x|^2 |\uparrow \rangle \langle \uparrow |_S + |y|^2 | \downarrow \rangle \langle \downarrow |_S$. Moreover, these two states are not just alternative descriptions or representations of different states of   knowledge, because they encode incompatible predictions for certain kinds of measurements. For example, if Bob is able to keep Alice and $S$ in a coherent state and prevent decoherence, he can in principle perform experiments which will detect the entanglement between Alice and $S$, whereas Alice herself will always find $S$ to be in a pure state. 

Currently we are not able to actually perform experiments of this kind, so we can't confirm experimentally that Bob really would see results in accordance with the state $\psi_{SA}$. Thus  one could always respond to this puzzle by  denying that   $\rho_{S}$ is really the right state to describe Bob's observations of $S$, as many  interpretations of quantum mechanics based on wavefunction collapse do \citep{Penrose1996,chalmers2021consciousness,pearle2024introduction}. But  if we want to maintain that unitary quantum mechanics  is `universal,'  in the sense that both Alice and Bob find that  it makes the right predictions when applied to their own observations, then $\rho_{S}$ must be the correct state for Bob to use   and either $| 0 \rangle$ or $| 1 \rangle$ must be the correct state for Alice to use, so it follows immediately   that  quantum states must sometimes be relativized, at least in an approximate empirical sense.
 
Therefore  putative solutions to the measurement problem which maintain the universality of unitary quantum mechanics typically tell us that quantum states can sometimes be relativized. Some of these solutions posit only  `effective' relationality \citep{adlam2025kindrelationalitydoesquantum}, meaning that there is always   an underlying non-relativized global quantum state but in cases like the Wigner's Friend scenario we can write down effective relativized states providing accurate characterizations of the results that will be obtained by specific observers - for example this occurs in the  Everett interpretation  \citep{Everett} and the de Broglie-Bohm model \citep{D_rr_1992, articlenorsen}. Other approaches posit `inherent' relationality \citep{adlam2025kindrelationalitydoesquantum}, meaning that   quantum descriptions must \emph{always} be relativized and the very idea of a non-relativized quantum state is incoherent - for example this occurs   in relational quantum mechanics \citep{1996cr}, various neo-Copenhagen approaches \citep{brukner2015quantum} and possibly QBism \citep{fuchs2023qbism}. 

What all these different  notions of quantum relationality have in common is  the following minimal understanding of what it means for a quantum state to be relativized  \citep{adlam2025kindrelationalitydoesquantum}:

\begin{quote}

\textbf{Relative quantum state:} The quantum state $\psi_S$ of system $S$ relative to system $F$ is the correct state to predict the outcomes of measurements performed by $F$ on $S$, but not necessarily the correct state to predict the outcomes of measurements performed by some other system $R$.

\end{quote}

 Relativization of quantum states in this specific sense is necessary and sufficient to address the apparent contradiction that arises if we try to maintain the universality of unitary quantum mechanics in the context of the Wigner's Friend scenario. 
 
 Note that this definition  deliberately does not specify what exactly it means for system $F$ to `measure' system $S$. In order to resolve the Wigner's Friend paradox of course the term must subsume  the measurements performed by Alice and Bob in that scenario, but the definition does not specify that $F$ must be a conscious observer, so the term `measurement' might refer to some rather general class of dynamical interactions which includes as a proper subset the kinds of interactions that we would normally describe as measurements. 
 
 Later in the article I will develop a concrete proposal about the nature of the relevant class of interactions, but for now the important point is simply that a measurement is a real physical process involving some kind of dynamical interaction between the measured system  and the system doing the measurement, so it is reasonable that the outcome of the measurement may sometimes depend on facts about the measuring system and its relation with the system undergoing measurement. It so happens that in classical physics we are able to set up measurements  in such a way that the result generally depends only on the nature of the measurement and the properties of the system being measured, but in quantum mechanics this is more challenging due to the holistic nature of the theory, and hence quantum states are sometimes relativized in the sense stated above.

\subsection{Relativization and Symmetries \label{GR}}

As \cite{Loveridge_2018} put it, `In classical physics, symmetry, reference frames and the relativity of physical quantities are intimately connected.' To see how this works, consider a physical theory and an automorphism $A$ of the class of mathematical models of that theory such that any two models related by $A$ represent possibilities which are not experimentally distinguishable. In this case we say that  $A$ is a symmetry of the theory \citep{greaves2011empiricalconsequencessymmetries}. Usually symmetries have a natural group structure, so we talk about the `symmetry group' $G$ of the theory - for example, global rotations are a symmetry group of all our  major physical theories.

 It is common to suggest that models connected by symmetry transformations represent states of affairs that are not really distinct \citep{Saunders2002-SAUPAL-2,wald2010general}, and I will assume here that this interpretation is correct. Thus if we have a total system $S \otimes F$ in isolation with no further system to which the joint state could be relativized, it follows that any two states of $S \otimes F$ connected by a symmetry transformation are physically identical, which means that any property of the joint system whose value may be changed by a symmetry transformation is not physically real in and of itself.
 
 Then consider some variable $V_S$ defined entirely on $S$  such that the value of $V_S$ can be changed by a transformation in $G$.  It follows that $V_S$ on its own is not physically real. However, suppose there is some variable $V_F$ defined entirely on $F$ such that there exists a   relationship $\mathscr{R}$ between  $V_S$  and  $V_F$  which remains the same under transformations in $G$   - for example, perhaps     the value of $V_S$   is always equal to the value of $V_F$. In this case the \emph{relationship}  $\mathscr{R}$ is physically real on its own, even though the individual values of $V_S$ and $V_F$ are not, so we can use this relationship to fix a value for $V_S$ \emph{relative} to a suitable reference state of $F$. For example, if it is an invariant fact that $V_F = V_S$ and we choose a reference state of $F$ where $V_F = 3$ then the value of $V_S$ relative to this reference state of $F$ is $3$. 

 A straightforward illustration of this is given by  a theory with a global rotation symmetry. Clearly in such a theory it is not meaningful to simply state an absolute orientation for an axis $F$, but since relative orientation is invariant under global orientations we can write down the orientation of $F$ relative to another axis $S$, using a reference state to fix the coordinate system. For example, if we choose reference state `up along the $Z$ axis' for $S$, then if $F$ is found to be parallel to $S$, the state of $F$ relative to $S$ will be written as `up along the $Z$ axis.' So in the context of symmetries, relativization arises in a natural way as a means of fixing values for symmetry-variant properties by  describing them relative to other physical systems. 

 \subsection{Relativization in General Relativity}

 In the context of general relativity, it is useful  to distinguish between local and global symmetries \citep{greaves2011empiricalconsequencessymmetries}. Global symmetries such as rotations have the feature that in order for them to be symmetries we must apply the same transformation everywhere. But  local symmetries allow us to vary the transformation arbitrarily from point to point. For example, diffeomorphisms are a local symmetry of general relativity \citep{wald2010general}. Performing a diffeomorphism  moves physical events, objects and quantities around on spacetime so they end up in  new spacetime locations, and therefore any fact pertaining to the content of a specific region of spacetime can be changed by a diffeomorphisms. Thus, if we assume that models connected by local symmetry transformations also represent states of affairs that are not really distinct, it follows that in a diffeomorphism-invariant theory such as general relativity, quantities assigned to specific spacetime locations are not physically real by themselves. 

 One might at first assume that fixing values by relativization as described in the previous section will work only for global symmetries, because in the local case we are always free to transform $F$ and $S$ separately so there could not be such an invariant relation  $\mathscr{R}$.  But in fact this   method can work for local symmetries too, because although local symmetries do not require that we implement the same transformation everywhere, nonetheless there are typically some restrictions on how local transformations can be combined  to produce valid models \citep{greaves2011empiricalconsequencessymmetries}, and sometimes these restrictions will ensure that a relation $\mathscr{R}$ is always preserved by valid symmetry transformations. 

 In the context of general relativity,  certain kinds of integrals involving the gravitational field are preserved by diffeomorphisms \citep{rovelli2022philosophical}, and thus we can still arrive at relational observables which arise from taking  highly non-local functions of the underlying variables and writing them in the form of a relation to a system which is, for the purpose of the relational description, treated as external. \cite{Rovelli_2002}  refers to the symmetry-variant quantities as `partial observables'  and the invariants as `complete observables.' That is,   partial observables are physical quantities `to which we can associate a measuring procedure leading to a number'  \citep{Rovelli_2002} such as spin, momentum, charge etc. By contrast, complete observables are typically fully relativized statements such as `the orientation of the spin of the particle relative to axis $A$ at the time when the clock $C$ reads $t$.' 
 
 Given the standard interpretation of symmetry transformations,  complete observables are physically real on their own, but partial observables are not.  This may initially seem puzzling, because many of the quantities that we observe in ordinary measurements, such as `the momentum of a certain particle at a certain time' are naturally interpreted as partial observables - and how could it possibly be that most of the things we actually measure are not physically real? The answer offered by \cite{Rovelli_2002} is that everything we measure is a partial observable but the \emph{results} of those measurements represent physically real complete observables, because the act of measurement itself relativizes the partial observable to our laboratory equipment. So let us now look in more detail at this claim about the relationship between observation and symmetry.

\subsection{ We only measure partial observables\label{partial}}

It is possible to find in the literature both  the claim that only invariant facts or complete observables are measurable,  and also  the claim that only partial observables are measurable. We find something like the former statement in \cite{Loveridge_2018}, which maintains that `observable quantities are invariant under symmetry.' We find something like the latter statement in Rovelli's definition of partial observables as physical quantities `to which we can associate a measuring procedure leading to a number' \citep{Rovelli_2002}. 

Despite superficial appearances,  these statements are not actually contradictory. Here it is helpful to employ the following simple schematic account of the process of measurement, which is intended as an elaboration on the account of  \cite{Rovelli_2002}.  When performing a measurement we of course do not make  direct mental contact with  intrinsic properties of  systems; measurement always involves an interaction between a variable $V$ and a measuring instrument which sets up  some kind of relation $\mathscr{R}$ between the variable and some appropriate variable $V'$ of the measuring instrument. For example, if $V$ is the orientation of a spin system, the result of measuring $V$ is not an abstract direction in absolute space but rather `the system is parallel to this axis' - that is, I might write down the result of the measurement as `up' but really that result is just a relativized way of expressing the invariant fact that the spin is parallel to the axis that I am using in my measurement. 

Given this schematization we can see why $V$ must always be a partial observable, because in order to measure $V$ we must be able to  directly or indirectly couple our measuring device to  $V$, and  only partial observables can  be directly interacted with  in such a way. As \cite{Rovelli_2014} puts it, `although we can interpret a given physical system in terms of its gauge-invariant observables, we nevertheless couple this same system with other systems using its non-gauge-invariant variables.' Here,  `gauge-invariant observables' correspond roughly to `complete observables,' while `non-gauge-invariant variables' correspond to   `partial observables,' so Rovelli is  arguing that partial observables  describe the ``handles" though which systems couple: we cannot interact directly with a complete observable because complete observables are by definition already complete, which means they do not have any free `handles' that we could couple our device to. 

In particular,   measurements  necessarily take place in time, so  in the process of making a measurement we are necessarily relativizing the quantity we are measuring at least to the  reading on our own laboratory clock at the time of the measurement, which will necessarily supersede parts of the relativization that originally defined the complete observable. For example, suppose I try to measure the complete observable given by `the relative distance between $A$ and $B$ at the time when clock $C$ reads $t$.' No matter what I do I will not succeed in measuring this specific observable, because instead I will end up measuring `the relative distance between $A$ and $B$ at the time when my laboratory clock $C'$ reads $t'$, ' where $t'$ reflects  the time at which I perform the measurement. That is, what I end up measuring is not the original complete observable but rather a partial observable, `the relative distance between $A$ and $B$' which gets relativized to my laboratory clock in such a way as to give rise to a new complete observable.

 With this schema in mind, Rovelli can be understood as emphasizing that the variable $V$ is always a partial observable, while Loveridge et al can be understood as emphasizing that the relation $\mathscr{R}$ corresponding to the \emph{result} of the measurement is always a complete observable, so the two statements are perfectly compatible. 
 
 The reasoning behind Rovelli's description of partial observables as variables `to which we can associate a measuring procedure' is that when a scientist describes what they are measuring, they don't normally include the measuring device and the laboratory clock in that description, and thus the thing we typically \emph{describe} ourselves as measuring   - e.g. `the momentum of the particle' - is the unrelativized partial observable $V$. Furthermore there are good reasons to talk this way,  for as noted by \cite{rovelli2022philosophical}, in the context of general relativity the invariant fact $\mathscr{R}$ will be highly nonlocal whereas the quantity $V$ is a local function of the metric, since it is located in the place determined by the measuring instrument. Thus it is convenient to describe ourselves as measuring $V$, since this admits a local description that would not always be available if we focused only on $\mathscr{R}$. 
 
 Whereas the reasoning behind the assertion of Loveridge et al that all observable quantities are invariant is that the \emph{result} of the measurement really gives us information not about the partial observable $V$ but about the underlying complete observable $\mathscr{R}$, so every possible observation must be linked to a complete observable in some way. As Loveridge et al explain, `in quantum mechanical laboratory experiments, the measured statistics pertain not to some absolute quantity, but rather to an observable, relative quantity, corresponding to the system and apparatus combined, along with the appropriate high localisation limit on the side of the apparatus.'

\section{Connecting Classical and Quantum Relativization \label{connect}} 

 Given that relativization in classical and relativistic physics is so tightly connected with symmetry, it is a little surprising that the standard   Wigner's Friend argument for relativization in quantum mechanics, as laid out in section \ref{quantum}, appears to have nothing to do with symmetries. Moreover, note that a number of formalisms developed to study quantum relationality do in fact rely heavily on methodologies related to symmetries, including  the quantum reference frame approach \citep{de_la_Hamette_2020,janssen2026quantumreferenceframesspacetime, Loveridge_2018} and the study of superselection rules \citep{Bartlett_2007}. This  suggests that it should be possible to make a closer connection between quantum relationality and the symmetry-based relationality appearing in classical and relativistic physics. 

It is sensible to begin such an enquiry with the simplest possibility: perhaps they are actually the same type of relationality. This would involve the hypothesis that all quantum states describe degrees of freedom which vary under some symmetry, so they need to be stated relative to a reference system for just the same reason as any symmetry-variant quantity needs to be stated relative to a reference system. Since this is the most straightforward way in which quantum relationality and relativistic relationality could be connected, we should establish whether it is a possibility before moving on to more complicated conjectures. So my goal in this article is to establish whether  this idea is compatible with our current understanding of quantum mechanics, or whether it is already ruled out by the existing evidence. 
 
I will begin in this section by discussing how this approach would work in the case of quantum states where it is obvious that the property described by the state varies under some global symmetry. In section \ref{variant} I will argue that this approach can plausibly be generalized to all quantum states.

\subsection{Quantum States of symmetry-variant properties}

Many of the quantum states that we deal with in real experiments  assign probabilities over the values of some property which known to vary under a global symmetry transformation. For example, this is true for quantum states of position (changed by global translations), and quantum states of spin (changed by global rotations), and probably also quantum states of energy (changed by global shifts in the standard of `zero energy') and charge (changed by global charge conjugation).  

To be concrete, in this section I will focus on quantum states of spin, though the points generalize straightforwardly to other global symmetries with a similar group structure. So suppose we write down a quantum state describing a spin-1/2 system:

\begin{equation} | \psi_S \rangle = \frac{1}{\sqrt{3}} | \uparrow \rangle + \sqrt{\frac{2}{3}}| \downarrow \rangle \label{spinstate} \end{equation}

 Here,  $| \uparrow \rangle$ is defined as the state in which the spin of the system points up in the Z direction, and $| \downarrow \rangle$ is defined as the state in which the spin of the system points down in the Z direction. But if global rotations are a symmetry of the theory within which we are working,  then the properties `up along the Z direction' and `down along the Z direction' have no physical significance on their own, so the state $\psi_S$ as written above could not possibly be an absolute quantum state: it can only be meaningful when defined relative to a reference system specifying a $Z$ axis.  

In a sense this is just the well-known fact that the mathematical representation of a quantum state depends on a choice of basis. However, in contexts where we presuppose an absolute background spacetime, it is natural to think  there is an observer-independent fact of the matter about the direction along which the particle's spin is really pointing, so different choices of basis just correspond to different possible representations of this underlying absolute spin. But if we take the global rotation symmetry of the theory seriously there cannot be a fact of the matter about the direction along which the spin is really pointing independent of a reference system used to fix a direction in space: assigning a direction to the spin is only meaningful when the assignation is relativized to an actual, physical reference axis. So once symmetries are taken into account, choosing a basis is no longer  merely an abstract decision   about how to represent the system - the choice of basis must be linked in some way  to an actual physical system used to fix  a direction in space. 

To see this, note that if  no suitable reference system is available then the effective state of the system is not $\psi_S$,  nor is it simply $\psi_S$ written in a different basis. Without a reference system some of the degrees of freedom cannot be operationally accessed and indeed have no real meaning, so the resulting state is not just a redescription of   $\psi_S$ but something else altogether. Specifically, the state of $S$ describing the degrees of freedom available to an external observer who has no orientation reference frame for $S$ is given by incoherent group averaging - i.e. averaging over all of the possible  transformations in the rotation group with respect to the  Haar measure $dg$ which is invariant under the group $G$ \citep{Bartlett_2007}:

\begin{equation} \psi_S \rightarrow \rho_S =  \int_G  U_S(g) | \psi_S \rangle \langle \psi_S | U_S(g)^{\dagger} dg \end{equation}

 The resulting averaged state is invariant under transformations in $G$. This operation produces a mixture over all eigenspaces of the unitary representation of the group, so the density matrix $\rho_S$ is block-diagonal, with each block corresponding to one of the eigenspaces \citep{Bartlett_2007}. 

Evidently  $\rho_S $ has many properties which are different to $\psi_S$, since it is a mixed state rather than a pure state -   it corresponds to an incoherent mixture of spins `smeared' in all possible directions, whereas the original state $\psi_S$ is an eigenstate of spin pointing in some particular direction. The fact that we get an entirely different type of state in the absence of a suitable reference system makes it clear that the relativization associated with the need to specify a reference system for spin is not merely a matter of making a choice of representation - a state like $\psi_S$ is simply not meaningful at all without relativization to a real physical reference system. 

So it is clear that a quantum state describing a symmetry-variant quantity such as spin must always  be relativized to a reference system: in a theory with a global rotation symmetry, it makes no sense to write down a quantum state such as $\psi_S$ which assigns  probabilities over different spin orientations without relativizing those probabilities to a reference axis. The statement that `$S$ has spin up relative to the axis defined by the reference system $R$,' depends on $R$ just as much as it depends on $S$, and thus a quantum state which predicts and systematizes the probabilities for finding the spin pointing in various directions relative to $R$ likewise   depends on $R$ just as much as it depends on $S$.

 \subsection{Quantum relativization \label{form}}

 Let us now consider why relativization of symmetry-variant quantities should lead  to the specific form of quantum relativization defined in section \ref{quantum}. 
 
 First, I will assume that part of what we mean when we talk about a system $F$ `measuring' another system $S$ is that, if a symmetry-variant property of the system $S$ is measured, then $F$  itself or some part of $F$ is used as a reference to which the state of $S$ is compared. For example, if a measuring instrument $F$ measures the spin of $S$, this means that the orientation of $S$ is being compared to some  axis which forms part of $F$. Thus the state $\psi_S^F$ of system $S$ relative to $F$ should be understood as the state which predicts the outcome of a measurement comparing the  orientation of $S$ to the relevant axis in $F$. The role of the reference state of $F$ is then simply to fix the coordinate system, so for example if the reference state of $F$ is  $| \uparrow_Z \rangle_F$ this means the outcome of the measurement is to be represented in a coordinate system such that the axis in $F$ is by definition pointing up along the $Z$ axis, e.g.  if $S$ is found to be parallel to the $F$ axis, that corresponds to the $| \uparrow_Z \rangle_S$ outcome in this coordinate system. Thus it is clear that the state $\psi_S^F$ is the right state to predict the outcomes for a measurement on $S$ performed by $F$, simply because this is the state written to fit the axis and coordinate system that will be used in such a measurement.
 
 But   $\psi_S^F$ may not be the correct state to predict the outcomes of  measurements performed on $S$ by some other system $R$, because a  measurement performed by $R$ will be comparing $S$ to an axis which forms part of $R$ instead, and in general the orientation of $S$ relative to another reference axis will be different from its orientation relative to the axis in $F$, so a quantum state  assigning probabilities over orientations relative to $F$ will generally have to be transformed in some way to arrive at appropriate predictions for $R$.

In particular, this form of relativization can plausibly address the puzzle of the Wigner's Friend scenario in the case where the initial state $\psi_S$ of the system $S$  measured by Alice is associated with a symmetry-variant property such as spin.  As described in more detail in   \cite{refframeloophole}, if it is assumed that global rotations are a symmetry of the theory then we can define two states  $| P \rangle_A$ and $|P' \rangle_A$ which both represent Alice as observing the particle to be parallel to her own axis, but which have different overall orientations relative to Bob. That is, these two states are identical with respect to facts such as relative orientation which belong to the invariant subspace for the rotation group, so they are indiscernible from within Alice's laboratory, but they stand in different external relations to Bob's axis.  Then we may suppose that after Alice has performed her measurement she ends up in a state like the following: 

\begin{equation} \psi_{SA}' =   x | \uparrow \rangle_S | P \rangle_A  + y | \downarrow \rangle_S | P' \rangle_A   \end{equation}

 In both branches of this state, $S$ has a well-defined spin orientation relative to Alice's axis, so $ \psi_{SA}'$ tells us that   $| \uparrow \rangle$ is the correct state for Alice to use for $S$ since this is the state that accurately describes the orientation of $S$ with respect to her internal laboratory axis. But the same time  the superposed state $\psi_{SA}'$ as shown above is the correct state for Bob to use for $S$ and $A$ together, since it is the state that accurately describes the orientation of $S$ together with the rest of Alice's laboratory relative to his own external axis.  In particular, it is true that relative to Alice  $S$ is in a pure state but also true that relative to Bob $S$ and $A$ are entangled \emph{in the orientation degree of freedom}; there is nothing mysterious about the fact that Alice and Bob assign different orientation states to $S$, because it is simply a consequence of the fact that orientation must always be stated relative to an axis and Alice and Bob are using different axes. Thus taking account of the role of reference systems in defining the quantum states of symmetry-variant properties can provide a straightforward way of explaining why Alice and Bob might need to assign different quantum states to the same system at the same time.

\subsection{Relativization with Symmetry Groups \label{loveridge}}

In order to generalize this approach we can appeal to a technical formalism  developed by \cite{Loveridge_2018}  describing how relativized quantum states for symmetry-variant quantities come about.  Consider a symmetry group $G$ with a unitary representation $U_S(g)$ on the system $S$. Loveridge et al  define a `relativization mapping' $\textyen$ which takes an operator $A$ on $\mathscr{H}_S$ to an operator $\textyen(A)$ on the joint Hilbert space $\mathscr{H}_S \otimes \mathscr{H}_F$ such that $\textyen(A)$ is invariant under transformations belonging to the  group $G$:

\[ \textyen(A) =  \int_G U_S(g) A \ U_S(g)^* \otimes F(dg) \] 

Here, $F()$ is simply a positive operator-valued measure on $\mathscr{H}_F$ which is covariant with respect to the representation $U(g)$ of the symmetry\footnote{Here a positive operator-valued measure is defined as a map from a sigma-algebra to the Hilbert space of the relevant system, and we say it is covariant if $F( g.X) = U(g) F(X) U(g)*$.}.  We say that a sequence of states $\phi_i \in \mathscr{H}_F$ is a `localising sequence' with respect to $F()$ at $X$ if it is the case that  $\lim_{i \rightarrow \infty}  \langle \phi_i | F(X) | \phi_i \rangle = 1$. That is, in the limit, the states $\phi_i$ are approximately eigenstates of $F(X)$.

 \cite{Loveridge_2018} show that if we have a sequence which is localising with respect to $F(X)$ when $X$ is the identity element of the group $G$, we get the following result: 

\[\lim_{i \rightarrow \infty} Tr((\rho_S \otimes |\phi_i \rangle \langle \phi_i |)\textyen(A)) = Tr(\rho_S A)\]

To understand this, suppose that our reference system $F$ has  a localising sequence as described above, and choose a reference state $| \phi_i \rangle$ of $F$ which has high localisation. Then suppose that the whole system $S \otimes F$ is in an invariant state which can be written in the form $GA(\rho_S \otimes |\phi_i \rangle \langle \phi_i |) $, where $GA()$ represents the group averaging operation defined in section \ref{connect}. It follows that for any measurement on $\mathscr{H}_S$, the probabilities we get from acting on the invariant joint state $GA(\rho_S \otimes |\phi_i \rangle \langle \phi_i |)$ with the  measurement $\textyen(A)$   are approximately the same as the probabilities we get from acting on the state $\rho_S$ using the  measurement $A$. That is, in this case the state $\rho_S$  gives approximately the same information as the joint state of $S$ and $F$, so it can be thought of as  just another way of conveying the same information as the joint state, and in that sense it is the appropriate state of $S$ relative to $F$ for the reference state $|\phi_i \rangle.$

It is important to this construction that the state of $S$ is relativized not merely to $F$ simpliciter but to a specific reference state of $F$. This can be understood from the schematic description in section \ref{GR}: if the relation $\mathscr{R}$ between the variable $V_S$ of $S$  and the variable  $V_F$ of $F$ is invariant (e.g.  $V_S$ is always equal to $V_F$) then we can arrive at a well-defined value of $V_S$ relative to some specific value of $V_F$, but it doesn't make sense to ask for the value of $V_S$ relative to $V_F$ or $F$ simpliciter.  

With this in mind, we can start from a definite, non-relative state of   $S \otimes F$ which corresponds to a  well-defined relation between $S$ and $F$ and yet nonetheless,  for certain reference states for $F$, it transpires that the state of $S$ relative to   $F$ is a coherent superposition of eigenstates of the corresponding variant quantity. As an example, consider  a composite  system $S \otimes F$ within a toy theory for which global rotations are the only symmetry group. The generators of the rotation group are the angular momentum operators $J_x, J_y, J_z$ associated with rotations around three orthogonal axes $x, y$ and $z$.  The eigenstates of $J_x$   are states such as  $| \uparrow_X \rangle$ and $| \downarrow_X \rangle$, associated with spin +1/2 and -1/2 respectively along the axis associated with $J_x$, and likewise for the other operators. 

Given the symmetry, all physically possible states of $S \otimes F$ must be invariant under global rotations, and this means that $S \otimes F$ cannot possibly be in a coherent superposition of different orientations. Moreover $S \otimes F$  will not even really be in a meaningful mixture of orientations, since the eigenstates of orientation are really all the same - the `mixture' is just a way of expressing mathematically the fact that all eigenstates of angular momentum represent the same physical state in this context. For example, one possibility is that the state of $S \otimes Y$ is an incoherent mixture of different states of the form $| \uparrow \rangle_S | \uparrow \rangle_F$ where $S$ and $F$ are parallel and pointing in different directions. What this represents is a state in which $S$ and $F$ have a well-defined relative orientation - they are parallel - but they do not point in any particular direction in space, since  directions are not meaningful without appeal to an external reference frame.

But what if we now consider the states of $S$ and $F$ individually?  \cite{Loveridge_2018} show that we may still see the emergence of `mutual coherence' -  that is, the state of the system $S$ relative to some reference state of  $F$ may be a coherent superposition of eigenstates of angular momentum, even though the state of  $S \otimes F$ does not involve any kind of coherent superposition. The coherence is mutual in the sense that the state $\rho_S$ of $S$ relative to a state $\rho_F$ of $F$ is coherent  if and only if $\rho_F$ is also coherent relative to $\rho_S$.  Therefore in this toy theory, we see that indeed, we can get relative states which are coherent even though the system as a whole does not exhibit any coherence, and this is a consequence of the fact that states describe a symmetry-variant property which  needs relativization to a reference system  and in particular a reference state in order to be meaningful.

\section{Variant versus invariant \label{variant}}

At this point we seem to encounter a puzzle. We have seen that there is a straightforward and natural reason why quantum states describing variant quantities might need to be relativized to a reference system. Yet the Wigner's Friend  argument commonly used to argue for the relativization of quantum states applies equally to any kind of quantum state, so insofar as the Wigner's Friend scenario is our motivation for thinking that quantum states are relativized, we seem to have  good reasons to believe that this is true for quantum states of any type of variable. What then should we say about quantum states of \emph{invariant} facts? 

Given the argument of section \ref{partial}, a straightforward answer to this question is available. If it is true that we only ever measure partial observables, then in all examples of quantum states that  have ever been experimentally demonstrated, the quantum degrees of freedom are partial observables. So we can perhaps just insist that  \emph{all}  quantum states pertain to  partial observables, which are symmetry-variant as long as diffeomorphisms are included in the class of relevant symmetries. Then we are free to maintain that all quantum states are relativized for the same reason, which is also the same reason that partial observables in general relativity must be relativized, i.e. because they have no physical meaning without being relativized to a reference system. Clearly if this could be made to work it would be a satisfying unification of quantum and relativistic relationality\footnote{ Note that this approach would imply that  quantum relationality is inherent and not effective: if all quantum states pertain to partial observables there cannot be any such thing as an  absolute quantum state, and hence the quantum states in a Wigner's Friend scenario are relational simply because all quantum states are relational, in virtue of the types of variables they describe.}.

There are also some strong conceptual motivations for such an approach. Given the discussion of section \ref{GR}, it is clear that partial observables are  conceptually and ontologically very different from complete observables, since the partial obsevables are ontologically dependent on the complete observables and not vice versa.  So  in some ways it would  be quite strange if partial and complete observables could both be captured in exactly the same mathematical formalism with the same dynamics, and thus it makes perfect sense that  quantum mechanics should apply only to partial observables and not complete observables.

Additionally, this approach provides a simple and intuitive  explanation for the prima facie puzzling phenomenon of quantum relativization. As noted in section \ref{form} the notion that Alice and Bob may assign different quantum states to the same system at the same time seems much less mysterious if we recognise that they are using these quantum states to characterize orientations  relative to distinct physical axes, and if \emph{all} quantum states describe symmetry-variant quantities, we can just say the same in general: quantum  relativization just comes from the use of different reference systems to describe the relevant variables, so there is nothing particularly puzzling about why different observers can end up with different states.   

  Of course, nothing about this story yet explains why the partial observables can be in superposition while the complete observables cannot, but we can make some preliminary observations. The key point about symmetry-variant quantities such as partial observables is that they are not physically real by themselves, and moreover we saw in section \ref{relationsproperties} that it is possible for the symmetry-variant quantities to be in superposition even when the corresponding invariant quantities are not in superposition. So in some sense a superposition of symmetry-variant quantities is not a superposition of anything `real' - if a quantum system is in a superposition of two different spin orientations, it is not the case that this system simultaneously has two different and incompatible intrinsic properties, but rather that a certain choice of reference system leads to relativized properties which look indefinite when expressed relative to a certain kind of reference state. With this picture in mind it seems quite natural that superposition should apply only to the `unreal' partial observables and not to the `real' complete observables, and this might potentially help to explain some  of the more counterintuitive features of quantum mechanics. 

This also  seems to be the interpretation favoured by \cite{Loveridge_2018} in their discussion of the relational formalism presented in section \ref{loveridge}: `The thesis of this paper is that  ... in quantum mechanical laboratory experiments, the measured statistics pertain not to some absolute quantity, but rather to an observable, relative quantity, corresponding to the system and apparatus combined, along with the appropriate high localisation limit on the side of the apparatus. This is quite general, and not specific to any particular absolute quantity, though in this paper special attention has been given to phase, angle and position.'

\subsection{Counterexamples\label{counterexamples}}

Against this, it will naturally be objected that we know for sure that there exist some quantum states associated with invariant facts. However,  once we take into account local symmetries such as diffeomorphisms it is not at all obvious that this is true, because the thesis that we only measure partial observables implies that in any example of a quantum state that has been experimentally demonstrated, the state must pertain to partial observables and so it is not in fact a counterexample to the thesis above.

To see how this applies to a specific example, consider the example of  the ground state wavefunction for the hydrogen atom \citep{10.1088/978-0-7503-1167-0ch12}: 

\[ \psi(r) = \frac{e^{r/a_o}}{\sqrt{\pi}a_0^{3/2}} \] 

This state is a coherent superposition of different values for the relative distance $r $ between the nucleus and the electron\footnote{Thank you to Anne-Catherine de la Hamette and Andrea di Biagio for this example}. One might at first think it is a counterexample to the idea that all quantum states describe symmetry-variant quantities, since  relative distance is invariant under many  important  symmetries, particularly  global symmetries such as translations, rotations and charge conjugation.  

But as a matter of fact, this depends on what we mean by relative distance. In general relativity there is indeed  an invariant definition of relative distance which involves taking appropriate integrals involving the gravitational field \citep{rovelli2022philosophical}. This kind of relative distance is a complete observable; I will refer to it as relative distance$_1$. But the relative distances we measure in the laboratory do not correspond in any straightforward way to such gravitational integrals, for when we measure relative distances in the laboratory what we  are really trying to learn is the relative distance \emph{at some time}; and  the time label assigned to a given value for the relative distance can be changed by performing a diffeomorphism, so ultimately the relative distance that we actually measure is a partial observable  relativized to our laboratory clock. I will refer to this   as relative distance$_2$. 

So relative distance$_1$ is indeed symmetry-invariant but cannot be directly measured since it has no `handle' for us to couple to, while relative distance$_2$ can be directly measured but it is not an invariant. Moreover, it is evident that the $r$ in the state $\psi(r)$ should be understood as referring to relative distance$_2$ rather than relative distance$_1$, since $\psi(r)$ assigns probabilities over the results we should expect to see if we measure the relative distance  at a given time. So the state $\psi(r)$ is not actually a counterexample to the suggestion that all quantum states describe symmetry-variant quantities; once we include diffeomorphisms in the class of relevant symmetries then it is indeed a quantum state characterizing  a symmetry-variant fact, like the other examples discussed here. 

And clearly  much the same will apply to any other putative counterexample, because in the end all quantum states that appear in  examples that can be demonstrated in experiments must pertain to measurable quantities, so, insofar as it is plausible that  we only measure partial observables, it is also plausible that we have only ever seen examples of  quantum states of partial obervables.

\subsection{Variant quantities and invariant relations \label{relationsproperties}}

Another type of worry arises from noting that any symmetry-variant fact whose value can be fixed by relativization corresponds to a relation which is invariant under the the relevant symmetry. For example, orientation is not invariant under global rotations, but relative orientation is invariant under global rotations. Thus one might initially think it is inconsistent to say that  variant facts can enter into coherent superpositions while  invariant facts cannot, because the state of a variant quantity of $S$ relative to $F$ is simply another way of expressing the invariant relation between $S$ and $F$, so if the state of a  variant quantity of $S$ relative to $F$ is superposed then the invariant relation between $S$ and $F$ must also be superposed.

However, using the framework of \cite{Loveridge_2018} as introduced in section \ref{loveridge}, we can see that this is not right.  Recall that  to derive a state of $S$ from a state of $S \otimes F$ we must choose a reference state for $F$, so we are not relativizing the state of $S$  merely to the \emph{system} $F$, but also to a particular state for it. And   we saw already in section in section \ref{loveridge} that due to the phenomenon of mutual coherence it is possible for the state of $S$ relative to a suitable reference state of $F$ to be a coherent superposition of the relevant symmetry-variant quantity, even though the state of $S \otimes F$ does not involve a coherent superposition of the corresponding invariant relation. So once the role of reference states is taken into account it is clear that superpositions of variant facts do not necessarily entail superpositions of the corresponding invariant relations. 

It should be emphasized that \cite{Loveridge_2018} do not deal explicitly with the case of diffeomorphism symmetry or other local symmetries, and as noted in section \ref{counterexamples}, in order to make this approach work in general it is probably necessary to include diffeomorphisms among the relevant symmetries. So there is still an open question about whether the same treatment can  work for the diffeomorphism group; but the existing framework is adequate to show that it is consistent in principle to imagine that variant facts can enter into quantum superposition while the corresponding invariant relations do not enter into quantum superposition.

\subsection{Going outside of quantum mechanics}
 
At this juncture one might object that the analysis of phenomena like the hydrogen atom in ordinary non-relativsitic quantum mechanics should not require us to appeal to relativistic concepts like diffeomorphism-invariance. But of course the discussion above does not imply that there is anything wrong with the usual analysis of the hydrogen atom! In general when we are studying the hydrogen atom or other non-relativistic quantum systems we are doing so relative to a well-defined time reference frame determined by our laboratory clock, so it is a perfectly good approximation to proceed as if we are working on  a fixed background spacetime; this is why we can usually disregard any issues relating to diffeomorphism invariance  in calculations pertaining to non-relativistic quantum systems. 

However, when we try to understand the nature of \emph{measurements} specifically, it is possible that a full understanding may require us to  take into account the fact that all of this is in the end relativized to a reference system. Such a conclusion should not be surprising, because it is well known that the phenomenon of measurement is  the aspect of quantum mechanics which most persistently eludes our conceptual grasp \citep{Maudlin1995-MAUTMP,WallaceIGUS,dowehaveany}, and in addition measurement is exactly the kind of process where the role of a reference system might become relevant. In particular, as discussed in \cite{adlam2026doeswavefunctioncollapserelational}, a measurement typically involves a system interacting with the very reference to which its state has been relativized, so it is not surprising that it becomes important in such a context to take into account all the relevant forms of relativization.

Moreover, although standard quantum mechanics disregards phenomena like diffeomorphism invariance, there are already results within the framework of ordinary quantum mechanics which indicate that many quantities commonly treated as `absolute' must in fact be understood as relative. For example,  the Wigner-Araki-Yanase (WAY) theorem \citep{osti_4866969,PhysRev.120.622,Loveridge_2011}  says that whenever we have an additive quantity $X$ which is conserved, it follows that for any variable $Y$ of the system $S$ that fails to commute with that quantity, we cannot measure the `absolute' value of $Y$ - assuming that measurements have to obey certain reasonable conditions characterizing what a measurement is supposed to achieve, it follows that we can only measure the value of $Y$ relative to the measuring apparatus. For example, because we expect momentum to be conserved and momentum does not commute with position,  we cannot measure the `absolute' position of a quantum system: `In the case of position measurements that obey momentum conservation, no WAY-type obstruction exists if one asks only for a measurement of the relative distance between the object and a “reference system”. In this case, when the reference system is provided by part of the apparatus, the measured observable can be given as the relative position' \citep{Loveridge_2011}.  So the WAY theorem serves as an indication that the need to relativize measured quantities is not completely  external to quantum mechanics  - it also receives some support from within the theory itself.

 \subsection{Beyond Experiment\label{counterexamples}}

If the argument above is accepted, we can be confident that the proposal  that quantum states always describe partial observables is consistent with all   direct experimental evidence for quantum mechanics. Still, one might think there are indirect theoretical reasons to believe that  quantum states can be used to describe complete observables too. For example, there are applications in cosmology and some versions of quantum gravity where it is quite common to write down a quantum state of `the universe as a whole' \citep{PhysRevD.28.2960,Bojowald_2015} and presumably such a state would have to pertain to complete observables, since there is nothing outside of the universe that it could be relativized to.

However, it should be emphasized that using quantum states in this way represents a significant extrapolation beyond the evidence we actually have. Quantum mechanics arose as an empirically-motivated systematization of observed measurement results, so it would not be very surprising if we were to find that it applies specifically to a special subclass of variables which subsumes all of the variables that we can directly measure. Of course it is not impossible that quantum mechanics should also apply to other classes of facts which are not measurable even in principle, but our current evidence does not provide any direct support for that claim, and thus it  should not be treated as the default to the exclusion of all other possibilities. 

Here it should be emphasized  that the idea that quantum mechanics has been formulated  to describe the kinds of quantities that can be measured  is not the same as saying that   quantum mechanics is purely operational. An operational approach to quantum mechanics would posit that the theory  only describes the outcomes of measurements and nothing beyond. By contrast, here we are using the category of `symmetry-variant' or `partial observables' to identify the types of variables to which the theory pertains, and then simply noting that as a matter of fact everything we can measure falls into this class -  `measurability' itself does not play a  role in defining any part of the content of the theory. In addition, this view is perfectly consistent with maintaining that quantum mechanics applies to all variables in the symmetry-variant category whether or not they are actually measured or able to be measured in some particular circumstances, so the approach does not rely on an operational stance and is perfectly consistent with robust realism about the content of quantum mechanics.

Now, one might be tempted to argue that if there exist quantum states of partial observables then we will inevitably end up with quantum states of complete observables too, since the linearity of quantum mechanics implies that superpositions can always be transferred from one set of variables to the other.   But as a matter of fact, we will  see in section \ref{measurement} that an examination of symmetry properties provides strong reasons to think that interactions between symmetry-variant and invariant quantities could not be linear in any case, so we should probably be sceptical about the existence of a  mechanism which would transfer superpositions linearly from partial observables onto complete observables, and thus it is perfectly reasonable to reject this kind of extrapolation.

 \section{The Measurement Problem \label{measurement}}

Thus far we have seen that  the idea that all quantum degrees of freedom are partial observables can  offer a straightforward way of connecting relativistic  relationality with quantum relationality. This suggestion clearly has consequences for the measurement problem, because it suggests that there does exist something akin to a `Heisenberg cut,' and that it falls exactly along the line between partial observables and complete observables. So in this section I will close by making some   observations about the status of measurement in this picture.

 \subsection{Linearity  \label{linearity}}
 
The Schr\"{o}dinger equation is linear, so if two state vectors are a solution to it,  any linear combination of those state vectors is also a solution \citep{Jordan_2009}. This means that if we know the Schr\"{o}dinger evolution of two states for a fixed period under a given Hamiltonian, $| \psi \rangle \rightarrow | \psi ' \rangle$ and $| \phi \rangle \rightarrow | \phi ' \rangle$, we also know that under the same circumstances the state $x | \psi \rangle  + y| \phi \rangle$ will evolve to $x| \psi' \rangle  + y | \phi' \rangle$. 
 
This linearity assumption plays an important role in setting up  the measurement problem \citep{Maudlin1995-MAUTMP}. That is, suppose an observer is measuring a quantum system in a coherent superposition of spin states, $|\psi_S \rangle = x | \uparrow \rangle_S +  y | \downarrow\rangle_S$, and prior to the measurement the observer is in some fixed `ready' state $| R \rangle_F$. Of course there is no hope of explicitly writing down the exact  state $| R \rangle_F$ since it involves an enormous number of degrees of freedom, and thus one might think there is no way to know what the post-measurement state of the observer and quantum system together looks like.

But here linearity  comes to our aid. Ordinary quantum theory tells us that if an observer correctly measures a system in the state $| \uparrow \rangle_S$, the observer will definitely obtain the outcome $\uparrow$ and the system will be left in the state $| \uparrow \rangle_S$, while the observer will be in the state $| see \  up \rangle_F $  which corresponds to `having seen the $\uparrow$ result.' The same goes  for $\downarrow$. So we know the measurement interaction must have the following effect on the basis elements: 

\begin{align*}
 |  \uparrow \rangle_S \otimes | R \rangle_F \rightarrow |  \uparrow \rangle_S \otimes  |  see \ up \rangle_F  \\
 | \downarrow \rangle_S \otimes | R \rangle_F \rightarrow |  \downarrow  \rangle_S \otimes |  see \ down  \rangle_F
\end{align*}

We can then invoke linearity and thereby, without knowing anything about the detailed quantum representation of the state $| R \rangle_F$, infer what the effect of the interaction must be in general:
\begin{equation}
\begin{aligned} \psi_{pre} =   x |  \uparrow \rangle_S \otimes  | R \rangle_F   + y |  \downarrow \rangle_S  \otimes  | R \rangle_F  \\ \rightarrow  \psi_{post} = x |  \uparrow \rangle_S \otimes  | see \  up \rangle_F   + y |  \downarrow \rangle_S  \otimes  | see \  down \rangle_F  \label{eq1} \end{aligned} \end{equation}

The way the linearity assumption is applied here presumes that coherent superpositions can always be  transferred between any pair of variables, as long as it is possible to set up an interaction between them.   

This application of linearity plays an important role in setting up the measurement problem \citep{Maudlin1995-MAUTMP}, for the final state $\psi_{post}$ that we have arrived at appears to represent the observer as observing two different outcomes at once, in apparent contradiction with the fact that we always observe exactly one outcome. Of course in practice to make predictions physicists simply suppose  that the wavefunction undergoes a stochastic `collapse' at some point during the measurement process,  breaking the linearity  and instantiating just one outcome for the measurement. But this cannot be the final solution to the measurement problem because it is intolerably vague and anthropocentric: if the stochastic collapse is a real physical process then we should be able to describe it in precise physical terms, and probably without making use of anthropocentric notions like `observers' or `consciousness.' Thus the problem we are faced with is that  we need to either find a clearer characterization of collapse, or explain clearly how to do without it.

\subsection{Transferring superpositions \label{transfer}}

Various existing collapse models approach this problem by postulating an extra dynamical mechanism which takes effect only in certain circumstances \citep{GRW,sep-qm-collapse,Penrose1996}. But the symmetry-based picture explored here points to a different way of understanding `collapse': rather than postulating an extra dynamical mechanism which interferes with the otherwise universally linear and unitary quantum evolution, we should simply maintain that the linearity of quantum mechanics does not apply to all possible types of couplings. And in  fact this is actually quite a natural conclusion, because once we take symmetries into account there seems to be something wrong with equation \ref{eq1} as given above.

Note that in this equation the  states $| see \ up \rangle_F, | see \ down \rangle_Y$ are supposed to be produced by a linear coupling which in effect just copies the information in the spin state of $S$ onto the observation state of the system $F$. If this is right, the states  $| see \ up \rangle_F, | see \ down \rangle_Y$ should have the same symmetry properties as the states $|  \uparrow \rangle_S , |  \downarrow \rangle_S $. But the spin orientation states $| \uparrow \rangle_S$ and $| \downarrow \rangle_S$ vary under global rotations, while the states  $| see \ up \rangle_F, | see \ down \rangle_Y$ do not vary under  global rotations, so the symmetry properties are not the same and thus  the $F$ states can't possibly be just a direct copy of the information in the $S$ states. Of course the $F$ states could instead be a direct copy of information about the \emph{relation} between $S$ and $F$, or about the relation between $S$ and some other reference system rather than just the variant $S$ states themselves, in which case the symmetry properties would indeed match - but in that case there is no longer  any  particular reason to expect the transition to be linear, so the equation \ref{eq1} still does not seem right. This also makes sense given the hypothesis that only partial observables and not complete observables are quantum degrees of freedom, for if that is the case then the right hand side of \ref{eq1} represents a superposition of variables that should not be able to be superposed, so it is clear that something about the transition must be incorrect.

 So a modification to standard unitary quantum mechanics seems inevitable:  linearity cannot be universal because variant-invariant couplings must involve something that looks like a stochastic `collapse.'  And of course  if it is true  that all measurable quantities are partial observables while all measurement  outcomes correspond to complete observables, then it follows that all measurements must involve some kind of variant-invariant coupling, and thus all measurements will necessarily involve some kind of non-linear collapse. So taking into account these symmetry properties seems to suggest a very natural explanation for the phenomenon of collapse during quantum measurements.

 \subsection{Comparison to other `collapse' views}

 Note that a collapse view of this kind would look quite different from typical dynamical collapse models. In particular, most collapse approaches \citep{GRW,sep-qm-collapse,Penrose1996} maintain that in the Wigner's Friend scenario Bob would not see results in accordance with the predictions of unitary quantum mechanics, because the \emph{entire} quantum wavefunction collapses when Alice does her measurements. But we saw in section \ref{form} that it is possible for the complete observables representing Alice's measurement outcome to instantiate a single well-defined outcome and yet for the symmetry-variant quantities corresponding to Alice's quantum degrees of freedom to remain in a superposition relative to Bob. Thus the symmetry-based approach allows that Alice will see a definite measurement outcome and yet Bob will still see results in accordance with the predictions of unitary quantum mechanics,  as long as he is measuring the appropriate symmetry-variant degrees of freedom. That is, because  this approach  posits a type of  collapse with a somewhat relational flavour, it allows us to maintain that quantum mechanics is equally valid for all observers and that it continues to work universally for everyone as long they apply it appropriately to partial observables. 

 Additionally, the nature of the collapse turns out to be rather distinctive in this picture. To get a clearer insight into this,  suppose we have described a set of systems $\{ S \}$ relative to a reference $R$. Thus when we write down the quantum state $\psi_S$ of the systems  $\{ S \}$, excluding the reference, that state is necessarily describing partial observables -  we get complete observables only when the reference $R$ is included. Ultimately only the complete observables are physically real, but when the  systems $\{ S\}$ interact or evolve amongst themselves this results in alterations to  the invariant relation between  the systems $\{ S \}$  and the reference $R$, so there are physically meaningful consequences at the level of the complete observables. Still,  as long as $R$ interacts only very weakly with the systems $\{ S \}$ it is a reasonable approximation to write the changes to the complete observables entirely in terms of changes in the variant quantities on the   $\{ S \}$ side: that is, under appropriate circumstances we can ignore  $F$ and describe the systems $\{S \}$ evolving and interacting linearly as if all the quantities involved are absolute.  

 But when  the reference $R$ interacts more strongly with a system $S'$ in $\{ S\}$, we can no longer neglect the  influence of $R$: in order to properly characterize what is happening at the level of the complete observables  we will need to say how the interaction with $R$ affects the system $S'$, and therefore we need to include $R$ itself in the description. Thus under these circumstances is no longer possible to express all physically relevant facts  in the form of  a linear evolution on the $\{ S \}$ side described relative to a static reference $R$; at this point the approximation fails and we have to give up the description of the systems $\{ S \}$ by themselves and then pick it up again after the system and reference cease to interact. So we get something that looks like a `collapse,' where the state of $\{ S \}$ undergoes a discontinuous transition during a time period where we are not able to describe $\{S \}$ relative to $R$.
 
 Thus in contrast to the instantaneous collapses appearing in most collapse models \citep{GRW,sep-qm-collapse,Penrose1996}, the symmetry-based approach would suggest that wavefunction collapse is probably not instantaneous.   Linear evolution breaks down once the system and reference begin to interact strongly, but as emphasized by \cite{adlam2026doeswavefunctioncollapserelational}, interaction is not a binary pointlike matter, and hence neither is collapse: what we refer to as a `collapse' simply corresponds to a period of time in which the approximation that allows us to model the systems $\{ S \}$ independently of the reference $R$ breaks down. In   order to fully develop a model of measurement and collapse in this scenario we would need to appeal to some formalism that allows us to move between the variant and invariant quantities and describe the coupling between them, and it is not obvious that we currently have any formal framework within which is suited to achieve this, but this would be an interesting area for further investigation.

\section{Conclusion}

Although it has often been suggested that quantum relativization has some kind of connection with the relativization featuring in general relativity, the nature of the connection has always been somewhat elusive. In this article I have investigated the possibility that they are  actually the same kind of relativization: quantum states must be relativized because the variables that they characterize are always symmetry-variant quantities.  Insofar as it is accepted that we only measure partial observables, this possiblity seems to be at least consistent with all existing evidence. And using the formalism of \cite{Loveridge_2018}, we can see that it is consistent in principle to suppose that only partial observables and not the corresponding complete observables enter into coherent quantum superpositions.

This approach also raises some  new possibilities for thinking about the measurement problem. In particular, the supposition that quantum mechanics applies only to partial observables would provide a straightforward reason why measurement outcomes do not enter into superpositions, and would potentially explain why the wavefunction must appear to `collapse' to produce a single definite outcome during the kinds of interactions that we refer to as measurements. This is of course not a complete solution to the measurement problem   because more work is needed to see how to characterize the coupling between variant and invariant quantities which results in such a nonlinear collapse, but the approach points to a potentially interesting new place to look for further insights into the nature of quantum measurement.  

 \section{Acknowledgements}

 Thanks to Leon Loveridge for very helpful comments on an earlier draft of this paper. This work was supported by the  John Templeton Foundation Grant ID 63209.


\begin{thebibliography}{}

\bibitem[Adlam, 2023]{dowehaveany}
Adlam, E. (2023).
\newblock Do we have any viable solution to the measurement problem?
\newblock {\em Foundations of Physics}, 53(44).

\bibitem[Adlam, 2025]{adlam2025kindrelationalitydoesquantum}
Adlam, E. (2025).
\newblock {What Kind of Relationality does Quantum Mechanics Exhibit?}
\newblock {\em European Journal for Philosophy of Science}, 16(8).

\bibitem[Adlam, 2026a]{refframeloophole}
Adlam, E. (2026a).
\newblock The reference frame loophole.
\newblock {\em arXiv}.

\bibitem[Adlam, 2026b]{adlam2026doeswavefunctioncollapserelational}
Adlam, E. (2026b).
\newblock Why does the wavefunction 'collapse' in relational approaches to quantum mechanics?

\bibitem[Araki and Yanase, 1960]{PhysRev.120.622}
Araki, H. and Yanase, M.~M. (1960).
\newblock Measurement of quantum mechanical operators.
\newblock {\em Phys. Rev.}, 120:622--626.

\bibitem[Bartlett et~al., 2007]{Bartlett_2007}
Bartlett, S.~D., Rudolph, T., and Spekkens, R.~W. (2007).
\newblock Reference frames, superselection rules, and quantum information.
\newblock {\em Reviews of Modern Physics}, 79(2):555–609.

\bibitem[Bojowald, 2015]{Bojowald_2015}
Bojowald, M. (2015).
\newblock Quantum cosmology: a review.
\newblock {\em Reports on Progress in Physics}, 78(2):023901.

\bibitem[Bong et~al., 2020]{Bong_2020}
Bong, K.-W., Utreras-Alarc{\'{o}}n, A., Ghafari, F., Liang, Y.-C., Tischler, N., Cavalcanti, E.~G., Pryde, G.~J., and Wiseman, H.~M. (2020).
\newblock {A strong no-go theorem on the Wigner's friend paradox}.
\newblock {\em Nature Physics}, 16(12):1199--1205.

\bibitem[Brukner, 2017]{brukner2015quantum}
Brukner, {\v{C}}. (2017).
\newblock {\em On the Quantum Measurement Problem}, pages 95--117.
\newblock Springer International Publishing.

\bibitem[Cavalcanti, 2021]{Cavalcanti_2021}
Cavalcanti, E.~G. (2021).
\newblock {The View from a Wigner Bubble}.
\newblock {\em Foundations of Physics}, 51(2).

\bibitem[Chalmers and McQueen, 2021]{chalmers2021consciousness}
Chalmers, D.~J. and McQueen, K.~J. (2021).
\newblock Consciousness and the collapse of the wave function.

\bibitem[de~la Hamette and Galley, 2020]{de_la_Hamette_2020}
de~la Hamette, A.-C. and Galley, T.~D. (2020).
\newblock Quantum reference frames for general symmetry groups.
\newblock {\em Quantum}, 4:367.

\bibitem[Di~Biagio and Rovelli, 2022]{pittphilsci19664}
Di~Biagio, A. and Rovelli, C. (2022).
\newblock {Relational Quantum Mechanics is about Facts, not States: A reply to Pienaar and Brukner}.
\newblock {\em Foundations of Physics}, 62(52).

\bibitem[Dieks, 2018]{https://doi.org/10.48550/arxiv.1801.09307}
Dieks, D. (2018).
\newblock {Quantum Mechanics and Perspectivalism}.
\newblock {\em arXiv:1801.09307}.

\bibitem[Durr et~al., 1992]{D_rr_1992}
Durr, D., Goldstein, S., and Zanghi, N. (1992).
\newblock Quantum equilibrium and the origin of absolute uncertainty.
\newblock {\em Journal of Statistical Physics}, 67(5-6):843–907.

\bibitem[{Everett}, 1957]{Everett}
{Everett}, H. (1957).
\newblock { `Relative State' Formulation of Quantum Mechanics}.
\newblock {\em Reviews of Modern Physics}, 29.

\bibitem[Fuchs, 2023]{fuchs2023qbism}
Fuchs, C.~A. (2023).
\newblock {QBism, Where Next?}

\bibitem[Ghirardi, 2016]{sep-qm-collapse}
Ghirardi, G. (2016).
\newblock Collapse theories.
\newblock In Zalta, E.~N., editor, {\em The Stanford Encyclopedia of Philosophy}. Metaphysics Research Lab, Stanford University, spring 2016 edition.

\bibitem[Ghirardi et~al., 1986]{GRW}
Ghirardi, G.~C., Rimini, A., and Weber, T. (1986).
\newblock Unified dynamics for microscopic and macroscopic systems.
\newblock {\em Phys. Rev. D}, 34:470--491.

\bibitem[Greaves and Wallace, 2014]{greaves2011empiricalconsequencessymmetries}
Greaves, H. and Wallace, D. (2014).
\newblock Empirical consequences of symmetries.
\newblock {\em The British Journal for the Philosophy of Science}, 65(1):59--89.

\bibitem[Greensite, 2017]{10.1088/978-0-7503-1167-0ch12}
Greensite, J. (2017).
\newblock The hydrogen atom.
\newblock In {\em An Introduction to Quantum Theory}, 2053-2563, pages 12--1 to 12--13. IOP Publishing.

\bibitem[Hartle and Hawking, 1983]{PhysRevD.28.2960}
Hartle, J.~B. and Hawking, S.~W. (1983).
\newblock Wave function of the universe.
\newblock {\em Phys. Rev. D}, 28:2960--2975.

\bibitem[Healey, 2022]{pittphilsci20397}
Healey, R. (2022).
\newblock Quantum relativity without relativism.
\newblock For publication in a special issue of Foundations of Physics on Relational Quantum Mechanics.

\bibitem[Janssen, 2026]{janssen2026quantumreferenceframesspacetime}
Janssen, D.~W. (2026).
\newblock Quantum reference frames for spacetime symmetries and large gauge transformations.

\bibitem[Jordan, 2009]{Jordan_2009}
Jordan, T.~F. (2009).
\newblock Why quantum dynamics is linear.
\newblock {\em Journal of Physics: Conference Series}, 196:012010.

\bibitem[Loveridge and Busch, 2011]{Loveridge_2011}
Loveridge, L. and Busch, P. (2011).
\newblock {`Measurement of quantum mechanical operators’ revisited}.
\newblock {\em The European Physical Journal D}, 62(2):297–307.

\bibitem[Loveridge et~al., 2018]{Loveridge_2018}
Loveridge, L., Miyadera, T., and Busch, P. (2018).
\newblock Symmetry, reference frames, and relational quantities in quantum mechanics.
\newblock {\em Foundations of Physics}, 48(2):135–198.

\bibitem[Maudlin, 1995]{Maudlin1995-MAUTMP}
Maudlin, T. (1995).
\newblock {Three Measurement Problems}.
\newblock {\em Topoi}, 14(1):7--15.

\bibitem[Morganti, 2009]{Morganti}
Morganti, M. (2009).
\newblock A new look at relational holism in quantum mechanics.
\newblock {\em Philosophy of Science}, 76(5):1027--1038.

\bibitem[Norsen, 2016]{articlenorsen}
Norsen, T. (2016).
\newblock Bohmian conditional wave functions (and the status of the quantum state).
\newblock {\em Journal of Physics: Conference Series}, 701:012003.

\bibitem[Oldofredi, 2021]{Oldofredi_2021}
Oldofredi, A. (2021).
\newblock {The Bundle Theory Approach to Relational Quantum Mechanics}.
\newblock {\em Foundations of Physics}, 51(1).

\bibitem[Ormrod and Barrett, 2022]{ormrod2022nogo}
Ormrod, N. and Barrett, J. (2022).
\newblock A no-go theorem for absolute observed events without inequalities or modal logic.

\bibitem[Pearle, 2024]{pearle2024introduction}
Pearle, P. (2024).
\newblock {\em Introduction to Dynamical Wave Function Collapse: Realism in Quantum Physics: Volume 1}.
\newblock Introduction to Dynamical Wave Function Collapse. Oxford University Press.

\bibitem[Penrose, 1996]{Penrose1996}
Penrose, R. (1996).
\newblock {On Gravity's role in Quantum State Reduction}.
\newblock {\em General Relativity and Gravitation}, 28(5):581--600.

\bibitem[Rovelli, 1996]{1996cr}
Rovelli, C. (1996).
\newblock Relational quantum mechanics.
\newblock {\em International Journal of Theoretical Physics}, 35(8):1637–1678.

\bibitem[Rovelli, 2002]{Rovelli_2002}
Rovelli, C. (2002).
\newblock Partial observables.
\newblock {\em Physical Review D}, 65(12).

\bibitem[Rovelli, 2014]{Rovelli_2014}
Rovelli, C. (2014).
\newblock Why gauge?
\newblock {\em Foundations of Physics}, 44(1):91–104.

\bibitem[Rovelli and Vidotto, 2023]{rovelli2022philosophical}
Rovelli, C. and Vidotto, F. (2023).
\newblock Philosophical foundations of loop quantum gravity.
\newblock In {\em Handbook of Quantum Gravity}, pages 1--28. Springer Nature Singapore, Singapore.

\bibitem[Saunders, 2002]{Saunders2002-SAUPAL-2}
Saunders, S. (2002).
\newblock {Physics and Leibniz's Principles}.
\newblock In Brading, K. and Castellani, E., editors, {\em Symmetries in Physics: Philosophical Reflections}, pages 289--307. Cambridge University Press.

\bibitem[Vidotto, 2022]{vidotto2022relationalontologycontemporaryphysics}
Vidotto, F. (2022).
\newblock The relational ontology of contemporary physics.
\newblock In {\em Quantum Mechanics and Fundamentality: Naturalizing Quantum Theory between Scientific Realism and Ontological Indeterminacy}, pages 163--173. Springer International Publishing.

\bibitem[Wald, 2010]{wald2010general}
Wald, R. (2010).
\newblock {\em General Relativity}.
\newblock University of Chicago Press.

\bibitem[Wallace, 2007]{WallaceIGUS}
Wallace, D. (2007).
\newblock The quantum measurement problem: State of play.
\newblock In Rickles, D., editor, {\em The Ashgate Companion to Contemporary Philosophy of Physics}. Ashgate.

\bibitem[Wigner, 1952]{osti_4866969}
Wigner, E.~P. (1952).
\newblock {Measurement of quantum-mechanical operations; Die Messun Quantenmechanischer Operatoren}.
\newblock {\em Z. Phys}, 133.

\bibitem[Wigner, 1960]{Wigner}
Wigner, E.~P. (1960).
\newblock The unreasonable effectiveness of mathematics in the natural sciences. {R}ichard {C}ourant lecture in mathematical sciences delivered at {N}ew {Y}ork {U}niversity, {M}ay 11, 1959.
\newblock {\em Communications on Pure and Applied Mathematics}, 13(1):1--14.

\end{thebibliography}
\end{document}